\documentstyle[12pt,aps,rotate]{revtex}
\input epsf
\newcommand{\frat}[2]{\frac{\textstyle #1}{\textstyle #2}}
\newcommand{\dmn}[2]{\mbox{$#1\!\cdot\! 10^{#2}\,$}}
\newcommand{\xmn}[2]{\mbox{$#1\!\times\! 10^{#2}\,$}}

\newcommand{\grpicture}[1]
{\epsfxsize=200pt
\hspace{5mm}
\parbox{200pt}{\epsfbox{#1.ps}}
\vspace{5mm}}
\begin{document}
\title{
\vspace*{-3cm}
\begin{flushright}
{\small
HUB-EP 97/97 \\
KANAZAWA 97-22 \\
ITEP-PH-/15-97 \\
revised version \\ 
}
\end{flushright}
\vspace*{0.5cm}
Semiclassical Approximation for Non-Abelian
Field \\ Strength Correlators in the Instanton Dilute Gas Model
\vspace*{1.0cm}
}
\author{
E.--M.~Ilgenfritz $^{1,3}$
\thanks{Supported by the Deutsche Forschungsgemeinschaft
under grant Mu932/1-4},
B.~V.~Martemyanov $^2$,
S.~V.~Molodtsov $^2$, \\
M.~M\"uller--Preussker $^1$,
and
Yu.~A.~Simonov $^2$ \\
$ $ \\
$^1$ {\it Institut f\"ur Physik, Humboldt--Universit\"at zu Berlin, Germany} \\
$^2$ {\it Institute of Theoretical and Experimental Physics, Moscow, Russia} \\
$^3$ {\it Institute for Theoretical Physics, University of Kanazawa, Japan} \\
}
\date{June 17, 1998}
\maketitle
\begin{abstract}
\par\vspace*{2.0cm}\noindent
Field strength correlators are semi-classically
evaluated in the dilute gas mo\-del of non-Abelian
sources (instantons) and compared with lattice data for
QCD at zero temperature. We show that one of the Euclidean invariant,
tensorial structures vanishes for configurations being
purely selfdual or anti-selfdual. We compute the invariant functions
contributing to the correlators within the two lowest orders in an
instanton density expansion. Fitting instanton size and density
for quenched and full QCD, we obtain a reasonable description.
\end{abstract}
\newpage

\section{Introduction}
\label{sect.1}

In the last years a systematic description of non-perturbative effects in QCD
has been given in terms of the gluon field-strength correlators
\cite{simdosch,sim,simshev}. They are of
immediate phenomenological importance in the model of stochastic confinement
of color charge, giving a detailed description of the level splitting of heavy
and light $\overline{Q}Q$ bound states, and in the description of
high-energy hadron and quark-(anti)quark scattering \cite{nachtmann}.
By now, there exist numerical results from lattice simulations concerning the
fundamental field strength correlators for pure gauge theory with gauge
groups $SU(2)$ \cite{campos,dig1,dDeb} and
$SU(3)$ \cite{dig2,dig3,dig4,Bali} over
physical distances ranging up to $O(1)$ fm. The correlators
have also been calculated near to the deconfinement transition in pure $SU(3)$
gauge theory \cite{dig3}. Very recently, this study has been extended to full
QCD  with four flavours of dynamical staggered quarks
\cite{dig4}. In this latter work, an interesting pattern of the quark mass
dependence both of the gluon condensate and of the gluonic correlation length
has emerged. The correlation length smoothly
changes from the lightest quark mass to the quenched results.
The extraction of the gluon condensate itself is afflicted with some
reservations about the renormalon ambiguity in the determination of the
Wilson coefficients of the operator product expansion. In Ref. \cite{dig4}
this problem is finally neglected. As demonstrated there, the
emerging mass dependence can be consistently
described by a low energy theorem which relates it to the zero-mass case
through the quark condensate.
The real problem hard to understand is the much stronger gluon condensate 
for pure Yang Mills theory.

In the direct computations of the gluonic field strength correlators
on the lattice the non-perturbative behaviour has been extracted with
a cooling procedure which serves the purpose to erase short-range perturbative
fluctuations. Then at various distances $d$ (in lattice step units)
gauge field structures are expected to become visible in the correlator if an
appropriate number of cooling iterations ($n \propto d^2$ due to the
diffusive nature of cooling) is applied.  In view of
these results it has become very likely that the field correlators have a
semiclassical origin and might provide information about the importance
and the physical parameters of those classical configurations
forming the basis of the semiclassical approximation.

Instantons - solutions of the Euclidean Yang Mills field equations -
are well-known examples of semiclassical configurations,
which exist both in the continuum and on the lattice.
Their contribution to the path integral quantization plays an
important role in explaining chiral symmetry breaking and
many phenomenological facts (see e.g.
\cite{tHooft,cal,mmpilgen,shur,diak,shurschaf,shur2}).

It seems quite in time to compute the field strength correlation
functions within a semiclassical model based on instantons
and to point out to what extent the latter are able to describe
the results of the lattice measurements mentioned above.
In principle, such a comparison should enable us
to quantify the importance of the semiclassical modes
forming the QCD vacuum or finite-temperature Euclidean fields,
respectively, and to provide their relevant physical parameters.

Performing the analysis of field strength correlators with the simplest
instanton solutions we will restrict ourselves to the $~SU(2)$ case,
being aware that the embedding into a larger $~SU(N_c)$ gauge
group can be easily realized. Therefore, the parameters of
instanton densities and sizes we are going to determine from a
comparison with lattice data will always refer to the physical $~SU(3)$
case.

Earlier studies of field strength correlators due to instantons
can be found in  \cite{Ba,Do}. There the results for the single instanton
approximation were obtained in terms of two-dimensional integrals.
Here we pay particular attention to the path dependence of the
phase factors. This can only be done numerically which was beyond the
scope of these papers. Ref. \cite{Do} points out the insufficiency of the
single instanton approximation.
We study here also the next order terms in a density expansion.
With the numerically evaluated invariant functions,
we estimate instanton density and size.

\section{The Field Strength Correlator}
\label{sect.2}
\subsection{General Statements}
\label{sect.2.1}

The gauge invariant two-point correlators of the non-Abelian
field strength are defined as
\begin{equation}
\label{eq1}
D_{\mu\rho,\nu\sigma}(x_1-x_2)=
\langle 0|Tr\{G_{\mu\rho}(x_1) S(x_1,x_2)
G_{\nu\sigma}(x_2) S^{\dagger}(x_1,x_2)\} |0 \rangle ~,
\end{equation}
where $G_{\mu\rho}= T^a G^a_{\mu\rho}$ is the field strength
tensor and $S(x_1,x_2)$ is the Schwinger-line phase
operator, {\it i.e.} the parallel transporter necessary to
join the field-strength operators at points  $~x_1,~x_2~$
in order to respect gauge invariance.
$T^a$ denotes the generators of the gauge group $SU(N_c)$.
The most general form of the correlator compatible with Euclidean $O(4)$
invariance at zero temperature is, in the notation of Ref. \cite{simdosch},
\begin{eqnarray}
\label{eq2}
D_{\mu\rho,\nu\sigma}(x) & = &
\left(\delta_{\mu\nu}\delta_{\rho\sigma}
     -\delta_{\mu\sigma}\delta_{\rho\nu}\right)
\left(D(x^2) +D_1(x^2)\right)+
\epsilon_{\mu\rho\nu\sigma}~D_2(x^2)+
\nonumber\\ [-.2cm]
&&\\[-.25cm]
&&+\left(x_\mu x_\nu \delta_{\rho\sigma}-x_\mu x_\sigma\delta_{\rho\nu}
+ x_\rho x_\sigma \delta_{\mu\nu}-x_\rho x_\nu\delta_{\mu\sigma}\right)
\frat{\partial D_1(x^2)}{\partial x^2}~,  \nonumber
\end{eqnarray}
with $x=x_1-x_2$ and $~D(x^2)~$, $~D_1(x^2)~$, $~D_2(x^2)~$
representing invariant functions. The invariant $~D_2$-term has
been added for later use. It is only relevant for cases in which $~CP~$
symmetry is violated or a restriction to sectors with topological charge
$~Q_t~\ne~0~$ is possible.

\noindent
It has been shown that  $~D_1~$ does not contribute to the area law of
Wilson loops \cite{sim,simshev}.
In the perturbative regime both invariant functions
$~D~$ and $~D_1~$ behave like $~1/x^4~$. Only  $~D_1~$
receives a contribution from one-gluon exchange \cite{simdosch,simshev}.
On very general grounds the perturbative contribution
to $~D~$ (which appears at one loop and higher orders)
was recently shown to be cancelled by higher correlator
contributions \cite{sim2}.
Here, we shall not discuss the perturbative contributions in more
detail. Instead we will concentrate on the contribution from
semiclassical configurations.

We consider the correlator of Eq. (\ref{eq1})
as an average of the expression in brackets
over gauge field configurations
with a weight equal to $\exp(-S_E)$ ($S_E$ is the Euclidean action of
the gauge field) times the determinant of the fermionic
Dirac operator. We are going to compare with lattice simulation data
which are taken partly in full QCD, partly within the quenched
approximation when the fermionic backreaction on the weight is neglected.
In our estimates only density and size of instantons play a role which are
- in the case of full QCD -
influenced by the effect of the fermionic determinant.
The instanton solution of the Euclidean field equations is taken with its
algebraic form
in the evaluation of the field strength and of the Schwinger lines.
The resulting expressions are then contributing in leading order to the
functional average.

At this place let us make a  general statement concerning the
leading order contribution of pure (anti-)selfdual gauge field configurations
to the field strength correlators.
This property of the one-instanton
contribution was mentioned before in Ref. \cite{Do}.

\noindent{\bf Lemma.}
Pure (anti-)selfdual field configurations can contribute only to
$D$~and~ $D_2$, but do not contribute to $D_1$.

\medskip\noindent
In order to prove this we assume that the gauge potential $~A~$ satisfes
the (anti-)selfduality condition on the field strength
\begin{equation}
\tilde{G}_{\alpha\gamma}
\equiv \frat12 \varepsilon_{\alpha\gamma\mu\rho} G_{\mu\rho}
= \pm G_{\alpha\gamma}~.
\end{equation}
Then the duality transformation $~\frat12 \varepsilon_{\alpha\gamma\mu\rho}
\frat12 \varepsilon_{\beta\delta\nu\sigma}~$, acting on both  field
strengths in Eq. (\ref{eq2}), provides
\begin{eqnarray}
\label{eq46}
\langle Tr \tilde{G}_{\alpha\gamma}S \tilde{G}_{\beta\delta}S^\dagger
\rangle & = &
\left(D(x^2) +D_1(x^2)+
x^2\,\frat{\partial D_1(x^2)}{\partial x^2}\right)
\left(\delta_{\alpha\beta}\delta_{\gamma\delta}
-\delta_{\alpha\delta}\delta_{\beta\gamma}\right)
+\epsilon_{\alpha\gamma\beta\delta}~D_2(x^2)
\nonumber \\ [-.2cm]
&& \\[-.25cm]
& - &
\left(x_\alpha x_\beta \delta_{\gamma\delta}
-x_\alpha x_\delta
\delta_{\beta\gamma}
+ x_\gamma x_\delta \delta_{\alpha\beta}
-x_\gamma x_\beta
\delta_{\alpha\delta}\right)~\frat{\partial D_1(x^2)}{\partial x^2}~.\nonumber
\end{eqnarray}
The sign of the third term of the right hand side of Eq. (\ref{eq46})
is just opposite to the sign of the corresponding term in Eq. (\ref{eq2}).
The left-hand sides of (\ref{eq2}) and
(\ref{eq46}) are equal  due to (anti-)selfduality. Thus, we get
$~D_1=const~$. From the requirement that the correlators should vanish
as $~x^2 \to \infty~$ we arrive with $~D_1 \equiv 0~$, q.e.d.

\noindent{\bf Corollary.}
Consequently, a  nonzero $D_1$ may appear only due to perturbative
fluctuations and in higher order of a density expansion for
distinct semiclassical contributions (due to the non-linear
interference of pairs
of (anti-)instantons).

\subsection{Lattice Results}
\label{sect.2.2}

The field-strength correlators (\ref{eq2}) have been estimated from lattice
two-point functions for pure $SU(2)$ and $SU(3)$ gauge theory at $~T=0~$
and $~T \neq 0~$ as well as for full QCD \cite{campos,dig1,dig2,dig3,dig4}.
After applying the cooling method the non--perturbative contributions
to the correlators have been shown to fall off exponentially at
distances between $0.1$ and $1$ fm. Fits to the cooled Monte Carlo
data still exhibit a remnant of the perturbative tail $~\sim x^{-4}~$.

In the following we want to consider the zero temperature case which, 
in the lattice measurements, is described by the following two correlators:
\begin{eqnarray}
\label{par/perp}
D_{||}(x^2)& =& D(x^2) + D_1(x^2) + x^2 \frac{\partial D_1(x^2)}{\partial x^2}
\nonumber\\ [-.2cm]
\\ [-.25cm]
D_{\bot}(x^2) &=& D(x^2) + D_1(x^2) \, . 
\nonumber
\end{eqnarray}
We will refer to the best fits to pure $SU(3)$ gauge theory data obtained
for several bare lattice couplings in \cite{dig2}:
\begin{eqnarray}
\label{eq5}
D(x^2)& =& A \exp(-|x|/\lambda_A)+ \frat{a}{x^4} \exp(-|x|/\lambda_a)~,
\nonumber\\ [-.2cm]
\\ [-.25cm]
D_1(x^2) &=& B \exp(-|x|/\lambda_B)+ \frat{b}{x^4} \exp(-|x|/\lambda_b)~,
\nonumber
\end{eqnarray}
with
\begin{eqnarray}
\label{eq6}
\frat{A}{\Lambda^{4}_L} \approx \xmn{3.3}{8}~, \quad
\frat{B}{\Lambda^{4}_L} \approx \xmn{0.7}{8}~, &\quad&
a \approx  0.69~,  \quad  b \approx  0.46~, \nonumber \\
\lambda_A = \lambda_B \approx  \frat{1}{\Lambda_L}\frat{1}{182}~,  &\quad&
\lambda_a = \lambda_b \approx  \frat{1}{\Lambda_L}\frat{1}{94}~.
\end{eqnarray}
In physical units, obtained from string tension data, this amounts to
$\lambda_A \approx 0.22$ fm, $\lambda_a \approx 0.43$ fm. The
fit was achieved with a very reasonable $~\chi^2 / N_{d.f.} \simeq 1.7~$.

There is another solution reported in \cite{dig2} with
\begin{eqnarray}
\label{eq6b}
\frat{A}{\Lambda^{4}_L} \approx \xmn{2.7}{8}~, \quad
\frat{B}{\Lambda^{4}_L} \approx 0.~, &\quad&
a \approx  0.4~,  \quad  b \approx  0.3~, \nonumber \\
\lambda_A = \lambda_B \approx  \frat{1}{\Lambda_L}\frat{1}{183}~,  &\quad&
\lambda_a = \lambda_b \rightarrow \infty~.
\end{eqnarray}
and a similar $~\chi^2~$ value.

The results show the invariant function $~D_1(x^2)~$ to be by a factor
$~O(5)~$ smaller than $~D(x^2)~$. This indicates that $~D(x^2)~$ is mostly
of non-perturbative nature.
For the perturbative contributions alone
the opposite relation $D_1>D$ has been proven to
be valid \cite{eidem}.

Within full QCD the same field strength correlators have
been measured for $N_f=4$ flavours of staggered fermions with the Hybrid
Monte Carlo (HMC) method \cite{dig4}. Results were obtained for two
quark masses $m_q=0.01,~0.02$ (in lattice units). Fits with the same ansatz
(\ref{eq5}) and (\ref{eq6}) have provided, for the case $m_q=0.01$,
the following parameters
\begin{eqnarray}
\label{eq6a}
\frat{A}{\Lambda^{4}_F} \approx \xmn{1.74}{10}~, \quad
\frat{B}{\Lambda^{4}_F} \approx \xmn{0.2}{10}~, &\quad&
a \approx  0.71~,  \quad  b \approx  0.45~, \nonumber \\
\lambda_A = \lambda_B \approx  \frat{1}{\Lambda_F}\frat{1}{544}~, &\quad&
\lambda_a = \lambda_b \approx  \frat{1}{\Lambda_F}\frat{1}{42}
\end{eqnarray}
with a  $~\chi^2 / N_{d.f.} \simeq 0.5~$. Determining the scale
parameter $\Lambda_F$ for full QCD from the estimate of the
$\rho$-meson mass one has $\lambda_A \approx 0.34$ fm, which is somewhat
larger than the quenched value quoted above. For the other quark mass value
$\lambda_A \approx 0.29$ fm was obtained. We conclude that qualitatively
the functions $~D$ and $~D_1$ behave similar in all cases considered.

In a recent paper \cite{Bali} the field strength correlators have been
investigated in the presence of a static $~q \bar{q}~$ pair
for quenched QCD (see also \cite{dDeb}).
The exponential decay of the form factors $~D$ and $~D_1$ has been seen in
this case even without cooling. The correlation lengths estimated are
compatible with the results mentioned before.

\subsection{Semiclassical Approximation}
\label{sect.2.3}

Next we want to work out the lowest order semiclassical approximation for
the field-strength correlator by separating the Gaussian integral over
quantum fluctuations on top of a single classical background field
from the gauge invariant product of field strengths simply evaluated
for that classical field. What then remains are zero-mode
integrations over the appropriate
set of collective coordinates $~\Gamma~$ characterizing
the classical field with a density function $~{\cal M}(\Gamma)$,

\begin{equation}
\label{Okt_25_1}
D_{\mu\rho,\nu\sigma}(x)=
\frat{1}{Z}\int d\Gamma {\cal M}(\Gamma) \cdot
Tr\{G_{\mu\rho}(x_1;\Gamma) S(x_1,x_2;\Gamma)
G_{\nu\sigma}(x_2;\Gamma) S^{\dagger}(x_2;\Gamma)\},
\end{equation}
where $G_{\mu\nu}(x;\Gamma)$ is the field strength tensor corresponding to
configurations $A_\mu(x,\Gamma)$.  To be more specific,
we imagine a model of the vacuum state that is semi-classically represented by
superpositions of $~N~$ instantons  and $~{\bar N}~$ anti-instantons
\cite{cal}

\begin{equation}
\label{Okt_25_2}
A_\mu(x,\Gamma)=\sum^{N}_{i=1}a_\mu(x;\gamma_i)+\sum^{\bar N}_{j=1}\bar
a_\mu(x;\bar \gamma_j)~.
\end{equation}
The $\gamma_i$ ($~{\bar \gamma}_j~$) denote the collective
coordinates of the $i$-th instanton  ($j$-th anti-instanton), which include the
position $z_i$, the group space orientation $\omega_i$,
and the size $\rho_i$.
The integration measure in Eq. (\ref{Okt_25_1}) is then expressed
by
$$
d\Gamma~=~\prod^{N}_{i=1} d \gamma_i
\prod^{\bar N}_{j=1} d \bar\gamma_j~, \qquad
d \gamma_i~=~d^4 z_i d \omega_i d \rho_i~, \qquad
d \bar\gamma_j~=~d^4 {\bar z}_j d \bar\omega_j d \bar\rho_j~.
$$
For practical use we will consider here only (anti-)instantons of fixed size.
This corresponds to the instanton liquid model invented in \cite{shur}
with a delta like size distribution. In principle,
a more realistic $\rho$-distribution with a selfconsistent exponential
infra-red cutoff (allowing to satisfy low-energy theorems) can be
obtained from the assumption that (anti-)instantons repel each other at
short distances \cite{mmpilgen,diak}.

If the instanton liquid or gas is sufficiently dilute we can approximate
the functional integral by an expansion in powers of the (anti-)instanton
densities
$~n_4~=N/V$ ($~{\bar n_4={\bar N}/V}~$).

Strictly speaking, the superposition ansatz (\ref{Okt_25_2})
makes sense as an approximate saddle point of the action only if the vector
potentials $a_\mu,$ ${\bar a}_\mu$ decrease fast enough.
This happens when the singular gauge expression is used for the
(anti-) instanton solutions $a_\mu,$ ${\bar a}_\mu$, instead of
the regular gauge form \cite{cal}.
The existence of a systematic expansion in higher order
contributions to the measure ${\cal M}(\Gamma)$ has been proven 
in Ref. \cite{levine}.
It will be pointed out in the next section that,
as long as one remains within the single instanton approximation,
the actual choice between an instanton in the regular or singular gauge
does not matter if the field strength correlator is evaluated on this
background with a straight line Schwinger phase factor inserted.
In the calculation of interference terms containing
$a_\mu(x;\gamma)$ and ${\bar a}_\mu(x;\bar\gamma)$, however,
the choice of the singular
gauge form of solution is essential, and will be used in section \ref{sect.4}.

\section{Field Strength Correlator in the One-Instanton Approximation}
\label{sect.3}

Let us compute next the explicit single-instanton
contribution to the field strength correlators (\ref{eq1}).
The leading term is given by the sum of
single instanton ($I$) and anti-instanton (${\bar I}$) contributions
\begin{eqnarray}
\label{eq8} D^{I}_{\mu\rho,\nu\sigma}(x_1,x_2)
& = & D^{I}_{\mu\rho,\nu\sigma} + D^{\bar I}_{\mu\rho,\nu\sigma}
\nonumber \\
& = & n_4 \int d^4 z~
Tr \left(G_{\mu\rho}(x_1;\gamma) S(x_1,x_2;\gamma) G_{\nu\sigma}(x_2;\gamma)
S^{\dagger}(x_1,x_2;\gamma)\right)+ \\
&& + (n_4,\gamma \to \bar n_4,\bar\gamma)~.
\nonumber
\end{eqnarray}
The integration over the (global) group orientation is trivial in this
case and has been omitted. We emphasize, that the Schwinger line phase
operator
is a path dependent matrix in the fundamental representation
\begin{equation}
\label{eq9}
S(x_1,x_2;z)=P\;\exp\left( i \int^{1}_{0}d t\, \dot{x}_\mu(t)
a_\mu(x(t);z)\right)~,
\end{equation}
where the vector potential $a_\mu=  T^a
a^a_{\mu}$,
in leading order of an expansion in the density,
belongs to the single instanton source localized at $z$.
Together with its adjoint, $~S^{\dagger}~$, $~S~$ takes care of the
parallel transport of the field strength tensor $G_{\mu\rho}$ from one
point of the measurement to the other.

We begin with the instanton solution in the so--called regular
gauge. Its SU(2) vector potential is expressed with $y=x-z$ as follows
\begin{eqnarray}
\label{eq51/52}
a_\mu^{a}(x;z) & = & 2 \,\eta_{a \mu \nu}
\frat{y_\nu}{y^2+\rho^2}~,~a=1,2,3, \\
G_{\mu \nu}^{a}(x;z) & = &
-4 \,\eta_{a \mu \nu}\,\frat{\rho^2}{(y^2+\rho^2)^2}~,~\mu,\nu=1,2,3,4~.
\end{eqnarray}
The rotational degrees of freedom in the group space have been omitted.
The t'Hooft tensor $~\eta_{a \mu \nu}~$ (and $~\bar{\eta}_{a \mu \nu}~$
for the anti-instanton) is defined in \cite{tHooft}.
The Schwinger line phase factor depends on a particular path
between the points of measurement $x_1$ and $x_2$.
For a straight line path between $x_1$ and $x_2$,
the Schwinger line can be written explicitely
\begin{equation} \label{eq53} S(x_1,x_2;z)=P\;\exp\left( i\, \tau^a
\eta_{a\mu\nu} \int^{1}_{0}d t\, \frat{(x_2-x_1)_\mu
y_\nu}{\rho^2+y^2}\right)~,
\end{equation}
where $\tau_a$ are the Pauli matrices and $y=x_1+(x_2-x_1)\,t-z$ denotes
a point running from $x_1$ to $x_2$ relative to the instanton center
$z$. The Schwinger line can be conveniently parametrized with $z_1=x_1-z$
and $z_2=x_2-z$ as follows
\begin{equation}
\label{eq54}
S(x_1,x_2;z)=\exp\left(\frat{i}{2} \tau^a \, n_a \Theta\right)\, .
\end{equation}
where
\begin{equation}
\label{eq55} \Theta= 2 \left(
z_1^{2}~z_2^{2}-(z_1 z_2)^{2} \right)^{1/2} \Psi ~,~~
~\Psi=\frat{\arctan (\chi_2)-\arctan (\chi_1)}
{\left( (z_1-z_2)^2 ~\rho^2+ z_1^{2}~z_2^{2}-(z_1 z_2)^{2}\right)^{1/2}}~,
\end{equation}
and
\begin{equation}
\label{eq57}
\chi_i=\frat{z_i(z_2-z_1)}
{\left((z_1-z_2)^2 ~\rho^2+ z_1^{2}~z_2^{2}-(z_1 z_2)^{2}\right)^{1/2}} ~,
\qquad i=1,2 ~.
\end{equation}
Here $n_a$ denotes the components of a unit vector in isotopic space,
\begin{equation}
\label{eq58}
n_a=\eta_{a\mu\nu}\frat{z_{2 \mu} ~z_{1 \nu}} {\left( z_1^{2}~z_2^{2}-
(z_1 z_2)^{2}\right)^{1/2}} ~.
\end{equation}
In what follows we shall compare the straigth line case (\ref{eq54})--
(\ref{eq58}) with a special case when the
Schwinger (path dependent) phase factor becomes trivial as $S(x_1,x_2;z)=1$
for all points $x_1$ and $x_2$.
This happens if the path connects the points $x_1$ and $x_2$ via the
instanton center $z$ along the two radial rays. In this case it is very
simple to perform the
integration over $z$ in (\ref{eq8}). The result can be expressed
by the following function of $x/\rho=|x_1-x_2|/\rho$
\begin{equation}
\label{eq59}
I_r\left(\frac{x}{\rho}\right)=\int d^4 z
\frat{\rho^4}{(z_1^2+\rho^2)^2(z_2^2+\rho^2)^2} ~.
\end{equation}
The result of the integration in (\ref{eq8}) for an instanton is
\begin{eqnarray}
\label{eq60}
D_{\mu\rho,\nu\sigma}^{I}(x)  =
( \delta_{\mu\nu}\delta_{\rho\sigma}-
\delta_{\mu\sigma}\delta_{\rho\nu}+\varepsilon_{\mu\rho\nu\sigma})
\,8~n_4~I_r\left(\frac{x}{\rho}\right) ~,
\end{eqnarray}
where
\begin{eqnarray}
\label{eq61}
I_r\left(\frac{x}{\rho}\right)=\frat{\pi^2}{2} \,\frat{\rho^2}{x^2}
\left(\frat{x^4}{c}+
\left( 1+\frat{x^4}{c}\right)
\frat{\rho^2}{\sqrt c}
\ln \left( 1+\frat{\sqrt c}{\rho^2} \,
\frat{\sqrt c+x^2}{\sqrt c-x^2} \right)-1 \right)
\end{eqnarray}
and $c=x^4 + 4\rho^2 x^2$, which coincides with the result obtained
in \cite{Do}.
The asymptotic behaviour of this function is
\begin{equation}
\label{eq62}
\lim_{x \to 0}  I_r\left(\frac{x}{\rho}\right) \to \frat{\pi^2}{6}~, \qquad
\lim_{x \to \infty} I_r\left(\frac{x}{\rho}\right)
\to \frat{4 \pi^2~\rho^4}{x^4}\ln \frat{x}{\rho} ~.
\end{equation}
The correlator due to an anti-instanton
is obtained by the replacement $\eta \to \bar {\eta}$ which results in
\begin{equation}
\label{eq63}
D^{\bar I}_{\mu\rho,\nu\sigma}(x)=
\left(\delta_{\mu\nu}\delta_{\rho\sigma}-
\delta_{\mu\sigma}\delta_{\rho\nu}-\varepsilon_{\mu\rho\nu\sigma}\right)
~8~\bar n_4 I_r\left(\frac{x}{\bar \rho}\right)~.
\end{equation}
In the instanton-anti-instanton dilute gas the observed
correlator in the lowest order of the density expansion
is the sum of (\ref{eq60}) and (\ref{eq63}).  Under the standard
assumption that the instanton and anti-instanton scale sizes are equal
we obtain for the invariant functions
\begin{equation}
\label{rep}
D(x^2)~=~8~ (n_4 + {\bar n}_4)I_r\left(\frac{x}{\rho}\right), \qquad
D_1(x^2)~=~0, \qquad
D_2(x^2)~=~8~ (n_4 - {\bar n}_4)I_r\left(\frac{x}{\rho}\right)\, .
\end{equation}
The vanishing of the one-instanton contribution to $~D_1~$ illustrates
the general theorem formulated
in section \ref{sect.2.1}.
Only in the case that the
densities $~n_4~$ and $~{\bar n}_4~$ are different, the $~\epsilon~$
tensor structure of the field strength correlator is different from
zero. In lattice computations one could select gauge field
configurations according to their net topological charge
$~Q_t~\ne~0~$. Such an ensemble would allow to extract $~D_2(x)~$.
However,
this is not the standard case. Therefore, we shall put $~n_4~=~{\bar n}_4~$
in the following.

If the Schwinger line is chosen to connect the points
$x_1$ and $x_2$ along a straight path
some corrections to formula (\ref{eq61}) appear.
One has to replace in (\ref{eq61})
$I_r(x/\rho)$ by $I_{ph}(x/\rho)$ where $I_{ph}(x/\rho)=
\frat{I_r(x/\rho)+2\,I_1(x/\rho)}{3}$
and
\begin{equation}
\label{eq64} I_1\left(\frac{x}{\rho}\right)=\int d^4 z
\cos\Theta \, \frat{\rho^4}{(z_1^2+\rho^2)^2~(z_2^2+\rho^2)^2} ~~,
\end{equation}
with $\Theta$ given in (\ref{eq55}).
The integral $I_1$ cannot be evaluated analytically.
Let us estimate the degree to what the field strength correlators
depend on the choice of
path in the Schwinger line. For the straight-line path between
the points $x_1,$ $x_2$ the result can be rewritten from
(\ref{eq59}), (\ref{eq64}) as
$$I_{ph}(x)=\left( \frat{1}{3}+\frat{2}{3}\langle
\cos \Theta\rangle\right) I_r(x)~,$$
where $\Theta$ is given in (\ref{eq55}) and the average is with respect to
the measure in (\ref{eq59}). An estimate for $\Theta$ can be given
for $x=x_1-x_2$,  with the instanton center located at distance $R$
(with $R \gg |x|$) from the midpoint $\frat{x_1+x_2}{2}$:
$\Theta \simeq \frat{2~ |x|~R}{R^2+\rho^2}$.
Thus the difference $1-\langle cos \Theta\rangle \approx
\frat{\langle \Theta^2\rangle}{2} \simeq
\langle \frat{2~x^2 R^2}{(R^2+\rho^2)^2}\rangle \simeq 0.2$, where
the typical instanton gas parameters $R \approx 1$ fm,
$x \approx \rho \approx 0.3$ fm have been inserted.
The result of the numerical integration for the straight line 
is compared with the expression     
(\ref{eq61}) for the radial path
gauge transporter in Fig. 1.
The correlator with the straight Schwinger line is roughly $20$ \%
smaller than for the $S=1$ case in accordance to the estimate,
and correspondingly the half width is smaller by roughly one third.
In the estimation of the (anti-)instanton density and size below,
the numerically calculated correlator with this minimal, straight
path Schwinger line should be used.

Considering the so-called singular gauge instanton solution
\begin{eqnarray}
\label{eq51/52s}
a_\mu^{a}(x;z) & = & 2 \,\bar\eta_{a \mu \nu}
\frat{y_\nu~ \rho^2}{y^2~(y^2+\rho^2)}~,~a=1,2,3, \\
G_{\mu \nu}^{a}(x;z) & = &
-\frat{8~\rho^2}{(y^2+\rho^2)^2}
\left\{\frat12\bar\eta_{a \mu \nu}
+\bar\eta_{a \nu \kappa}\frat{y_\kappa y_\mu}{y^2}
-\bar\eta_{a \mu \kappa}\frat{y_\kappa y_\nu}{y^2}
\right\}~,~\mu,\nu=1,2,3,4~,
\end{eqnarray}
we notice that it is related to the regular gauge (anti-) instanton
by a singular gauge transformation. With the straight Schwinger line
inserted, the correlator is identical to the correlator numerically
evaluated for the regular gauge instanton. Just in order to see how
the choice of the Schwinger line influences the field strength correlator, 
we can consider also a path for which 
the Schwinger line non-Abelian phase factor is equal to $S=1.$ This is
the path leaving $x_1$ along a radial ray
(starting from the instanton center) towards infinity and approaching  
$x_2$ from infinity along another radial ray, with an arc at infinity
in between.

The result of the integration over instanton position can be expressed
as a function of $x/\rho$ by the replacement $I~\to~I_s$ 
which is defined as
\begin{equation}
\label{eq59s}
I_s\left(\frac{x}{\rho}\right)=\int d^4 z~
\frat{\rho^4 \left(4~\frat{(z_1 z_2)^2}{z_1^{2} z_2^{2}}-1\right)}
{(z_1^2+\rho^2)^2(z_2^2+\rho^2)^2} .~
\end{equation}
There exists a closed expression of the integral 
(see the lower curve in Fig. 1)
\begin{eqnarray}
\label{60s}
I_s\left(\frac{x}{\rho}\right)=
\frat{\pi^2}{6}~\frat{\rho^2}{x^2}
\left\{-4~\frat{(x^2+\rho^2)(x^2+3\rho^2)}
{(x^2+4\rho^2)\rho^2}-2~\frat{x^6}{\rho^6}\ln\frat{x^2}{\rho^2}+
4~\left(1+\frat{x^2}{\rho^2}\right)^2\left(\frat{x^2}{\rho^2}-2\right)
\ln\left(1+\frat{x^2}{\rho^2}\right)\right.
\nonumber \\ [-.2cm]
\\[-.25cm]
+\left.
\left[\left(1+\frat{x^2}{\rho^2}\right)^2\frat{x^4}{\rho^4}-
3\left(1+\frat{x^2}{\rho^2}\right)\left(1+3\frat{x^2}{\rho^2}\right)-
\frat{3x^2}{x^2+4\rho^2}
\right]
\frat{\rho^2}{\sqrt{c}}\ln\frat{\rho^2(\sqrt{c}-x^2)}
{(\rho^2+x^2)\sqrt{c}+x^4+3\rho^2x^2}\right\}~.\nonumber
\end{eqnarray}
The asymptotic behaviour of this function is
\begin{equation}
\label{eq62s}
\lim_{x \to 0}  I_s\left(\frac{x}{\rho}\right) \to \frat{\pi^2}{6}~, \qquad
\lim_{x \to \infty} I_s\left(\frat{x}{\rho}\right)
\to \frat{2~\pi^2}{3}\frat{\rho^6}{x^6}~.
\end{equation}
Note that in the next to leading order (i.e. the approximation 
quadratic in the density)
only instantons written in the singular gauge can be
employed to form superpositions according to Eq. (\ref{Okt_25_2}).

The long-distance asymptotics of the field
strength correlator evaluated non-perturbatively
with single instanton contributions resembles a perturbative
contribution and does not show an exponential fall-off. This indicates
that only a strongly interacting instanton gas or liquid might
mimic the correct infrared behaviour of the theory. Nevertheless,
it is reasonable to ask, whether single instanton contributions
can describe reliably the behaviour of the field correlators at intermediate
distances, where lattice data are available. Independently of its concrete
form the instanton contributions to the correlation functions
should be compared only with the non-perturbative part of the lattice result,
{\it i.e.} with the pure exponential terms in Eq. (\ref{eq5}).

As already mentioned the lattice data for
$~D_1~$ are definitely smaller than those for $D$.
Due to the (anti-)selfduality, the tensor structure related to
$D_1$ even strictly vanishes in the leading order of the density expansion
in a dilute instanton gas picture.
Therefore, the leading instanton result points qualitatively into the
right direction.

The lattice measurements of the field strength correlators are
obtained with straight Schwinger line gauge transporters inserted
between the points $x_1$ and $x_2$. 
Comparing with the lattice data of Di Giacomo et al. \cite{dig2}
we can roughly estimate the instanton gas parameters.
Fitting the
instanton results for $~D~$ according
to Eq. (\ref{rep}) (with $~I_r~$,$~I_{ph}~$,$~I_s~$
related to different choices of the Schwinger line path)
to the first term of $D$ in (\ref{eq5}), {\it i.e.}
\begin{equation}
\label{estim1}
D^{inst}\approx A e^{-x/\lambda_A}~, \quad~D_1^{inst}\approx 0~
\end{equation}
within the range $1 < x/\rho < 5$,
we obtain
for $\rho$ and
$n^{t}_4=n_4+\bar{n}_4$, the radius and the total density of pseudo-particles,
the following results:
\\

\begin{center}
\begin{tabular}{|c|c|c|c|} 
\hline
quenched QCD                  &   $I_r$    &  $I_{ph}$  & $I_s$ \\ 
\hline
$\rho/\lambda_A$        & 0.78     &  1.35      & 2.12\\
$n^{t}_4~/\mbox{fm}^{-4}$& 6.19   &  4.03      & 2.29\\
\hline
$\rho~/\mbox{GeV}^{-1}$&0.87      &  1.51      & 2.36\\
$n^{t}_4~/\mbox{GeV}^4$&\dmn{9.39}{-3}&\dmn{6.11}{-3}&\dmn{3.47}{-3}\\
\hline
$n^{t}_4\rho^4$&\dmn{5.37}{-3}&\dmn{3.14}{-2}&\dmn{1.08}{-1}\\
$G_2= \frat{g^2}{4\pi^2}\langle (G_{\mu\nu})^2 \rangle$~
/$\mbox{GeV}^4$&\dmn{7.51}{-2}&\dmn{4.89}{-2}&\dmn{2.77}{-2} \\
\hline
\end{tabular}
\end{center}
where $G_2$ denotes the gluon condensate and $n^{t}_4\rho^4$ the
packing fraction.

Strictly speaking, only the column denoted by $~I_{ph}~$ with this function
evaluated in the instanton field according to the
straight Schwinger line
prescription should be compared with the result of
measurements obtained on the lattice. 
We
provide also the other estimates in order to get a feeling about the
sensitivity with respect to the choice of the Schwinger line path.

The difference to the value of the gluon condensate
$G_2=0.14(8) ~\mbox{GeV}^4$ extracted for the quenched theory
in Ref. \cite{dig3}, 
can be explained by the fact
that the latter result relies on the validity of the
exponential part of the fit down to $x=0$, while
the single-instanton correlator itself is flat for $x \to 0$.
The instanton size, practically identified with the correlation length,
is as usually adopted, $\rho \approx \frac13 ~\mbox{fm}$ within
the parametrizations leading to $I_{ph}$ and
$I_s$.  
The packing fraction $n^{t}_4 ~\rho^4$
almost coincides with the value for the instanton liquid.
Yet, the density estimated is bigger compared with the
density of $ 1 ~\mbox{fm}^{-4} $ usually adopted
in phenomenological applications.

However, the instanton liquid phenomenology makes references only to
full QCD with realistic quark masses.
Therefore, we give here a fit based on the instanton gas formula
to the full QCD lattice data, too, described by expression (\ref{eq5}) and
parameters (\ref{eq6a}).
We obtain from a fit in the range $1 < x/\rho < 5$ the following
parameter values:
\\

\begin{center}
\begin{tabular}{|c|c|c|c|} 
\hline
full  QCD            &   $I_r$    &  $I_{ph}$  & $I_s$ \\ 
\hline
$\rho/\lambda_A$        & 0.78     &  1.28      & 2.\\
$n^{t}_4~/\mbox{fm}^{-4}$& 0.72   &  0.55      & 0.64\\
\hline
$\rho~/\mbox{GeV}^{-1}$&1.34      &  2.21      & 3.45\\
$n^{t}_4~/\mbox{GeV}^4$&\dmn{1.09}{-3}&\dmn{8.34}{-4}&\dmn{9.7}{-4} \\
\hline
$n^{t}_4\rho^4$&\dmn{3.56}{-3}&\dmn{1.97}{-2}&\dmn{1.37}{-1}\\
\hline
\end{tabular}
\end{center}

\bigskip\noindent
Through an estimate based on the Banks-Casher formula
(see {\it e.g.}
the recent review by T.~Sch\"afer and E.~V.~Shuryak \cite{shurschaf})
\begin{equation}
\label{eq:casher}
\langle \overline{q} q \rangle =
- \frac{1}{\pi \rho} \sqrt{\frac{3 N_c}{2} n_4} \,\, ,
\end{equation}
we obtain 
for the quark condensate:
\\

\begin{center}
\begin{tabular}{|c|c|c|c|c|} \hline
full  QCD    &   $I_r$    &  $I_{ph}$  & $I_s$& experiment \\ \hline
$\langle \overline{q} q \rangle/~\mbox{GeV}^{3}$
&$-$\dmn{1.66}{-2}&$-$\dmn{8.82}{-3}&$-$\dmn{6.1}{-3}
&$-$\dmn{1.06}{-2}---~$-$\dmn{1.66}{-2}\\
\hline
\end{tabular}
\end{center}

We conclude that the nonperturbative tensor structure
$D(x^2)$  in the  field strength correlator at zero temperature
can be roughly described by a semiclassical picture based on
instanton-like non-perturbative field configurations. The extracted
parameters are in the expected ballpark when the fit is applied to 
the lattice measurements with dynamical fermions. This lends support to the
conjecture that (anti-)selfdual configurations are dominantly contributing to
the correlator and the gluon condensate as known from real QCD.

The gluon condensate in the quenched theory obtained in Ref. \cite{dig3},
however, 
is almost one order of magnitude bigger than that for full QCD with light
Kogut-Susskind quarks. With our instanton shape for the correlation function
we can reduce this to a roughly half as big estimate. Still this results in
an unexpectedly high instanton density. Due to the larger 
correlation length with dynamical light quarks, according to our fits, 
the packing fraction is bigger only by some $50$ \% in 
pure Yang Mills theory compared to QCD. In the following section we will study 
the influence on estimated instanton density 
and radius for the quenched case when corrections of second
order in the density to the field strength correlator are taken into
account.

\section{Second Order Density Contribution}
\label{sect.4}

In this section we shall present some estimates of the next order term in
a density expansion. We have to consider the field strength for a
superposition of solutions $a$ and $b$,
where both $a$ and $b$ can represent an instanton or anti-instanton,
\begin{eqnarray}
\label{Okt_25_3}
G_{\mu\rho}(a, b)&=&G_{\mu\rho}(a)+G_{\mu\rho}(b)+
\triangle G_{\mu\rho}(a, b)~,\nonumber\\ [-.2cm]
\\ [-.25cm]
\triangle G_{\mu\rho}(a, b)&=&-i  \{[a_\mu, b_\rho] +
[b_\mu,a_\rho] \}~.\nonumber
\end{eqnarray}
We neglect for the purpose of this estimate the Schwinger phase factors
which are known to give a $20$ \% effect in first order in the density
approximation. The interference between classical solutions
contributes terms of second order in the density, and the resulting field
strength is neither self\-dual nor anti-selfdual. Therefore we expect 
$~D_1~$ to receive the leading contributions in this order.
The functional weight (\ref{Okt_25_1}) specified for superpositions 
of Euclidean solutions is approximated by
an uncorrelated ansatz in terms of the single-(anti-)instanton density.
Therefore the second order contribution to the correlator has the form 
\begin{eqnarray}
\label{SecOrd}
D_{\mu\rho,\nu\sigma}^{(2)}(x_1,x_2) & = &
\frac{1}{2} \sum_{a,b=I{\bar I}}~n_4^{(a)}~n_4^{(b)}
\int d^4 z_1 \int d^4  z_2~
\int d \omega_1 \int d \omega_2~
\times \nonumber \\ [-.2cm]
&& \\ [-.25cm]
&& Tr \left\{ G_{\mu\rho}(a(x_1,\gamma_1),b(x_1,\gamma_2))~
G_{\nu\sigma}(a(x_2,\gamma_1),b(x_2,\gamma_2))\right\}. \nonumber
\end{eqnarray}

\medskip\noindent
When considering superpositions, one has to take the single (anti-)instanton 
field configurations $~a~$ and $~b~$ in  the singular gauge.
With the notation  $y=x-z$ their vector potential is expressed as follows
\begin{eqnarray}
\label{SecOrd_sng}
a_\mu^{a}(x;z)~ =~  \bar\eta_{a \mu \nu}~
y_\nu f(y,\rho)~,~b_\mu^{a}(x;z)~ =~ \omega_{aa'}\bar\eta_{a'\mu\nu}~
 y_\nu f( y,\rho)~,~~f(y,\rho)= \frat{2~\rho^2}{y^2(y^2+\rho^2)}~,
\end{eqnarray}
where $\omega_{aa'}$ denotes the relative color orientation of the pair
(replace $\bar \eta \to \eta$ for anti-instantons).

As was mentioned above, within the approximation linear in $n_4$
the correlator $D_1$  is strongly zero. However,  the
interference between (anti-)instantons  generates
non-trivial contributions to it. In general, one can easily derive
the contributions to the field
strength correlators  by averaging over the color orientation
of the (anti-)instanton $~a~$ relative to the (anti-)instanton $~b~$.
Only the $\triangle G_{\mu\rho}(a,b)~
\triangle G_{\nu\sigma}(a,b)$ contribution is non-zero after
this average has been taken.
Introducing the tensorial decomposition of the integral
\begin{equation}
\int d^4 y~ y_\mu (y+x)_\nu f(y,\rho) f(y+x,\rho)=\delta_{\mu \nu}
 J_1(x,\rho)+
x_\mu x_\nu J_2(x,\rho)
\end{equation}
in terms of invariant functions $J_1$ and $J_2$,
after some minor algebraic manipulations one can obtain
the following system of equations for the second order terms,
proportional to the number density of different {\it pairs} in the gas:
\begin{eqnarray}
\label{SecOrd_0}
D^{(2)}(x)+D^{(2)}_1(x)&=&\frat{N_c}{N_c^{2}-1}\frat{(n_4+\bar n_4)^2}{2}~
(3 J_1(x,\rho)+x^2 J_2(x,\rho))^2~,
\nonumber\\ [-.2cm]
\\ [-.25cm]
\frat{\partial D^{(2)}_1(x)}{\partial x^2}&=&-\frat{N_c}{N_c^{2}-1}
\frat{(n_4+\bar n_4)^2}{2}~
(3J_1(x,\rho)+x^2 J_2(x,\rho)) J_2(x,\rho).\nonumber
\end{eqnarray}
The integrals $J_1$ and $J_2$
can be computed analytically,
\begin{eqnarray}
\label{SecOrd_1}
J_1(x,\rho)&=&\frat{\pi^2\rho^2}{6}~\frat{\rho^4}{x^4}
\left\{4~\frat{x^2}{\rho^2}+2~\frat{x^6}{\rho^6}\ln\frat{x^2}{\rho^2}-
4~\left(1+\frat{x^2}{\rho^2}\right)^3\ln
\left(1+\frat{x^2}{\rho^2}\right)\right.
\nonumber \\ [-.2cm]
&&\\[-.25cm]
&-&\left.\frat{c^{3/2}}{\rho^6}\ln\frat{\rho^2(\sqrt{c}-x^2)}
{(\rho^2+x^2)\sqrt{c}+x^4+3\rho^2x^2}\right\}~.\nonumber
\end{eqnarray}
\begin{eqnarray}
\label{SecOrd_2}
J_2(x,\rho)&=&\frat{\pi^2}{3}~\frat{\rho^6}{x^6}
\left\{-8~\frat{x^2}{\rho^2}+2~\frat{x^6}{\rho^6}\ln\frat{x^2}{\rho^2}
\right.
-
4~\left(1+\frat{x^2}{\rho^2}\right)^2\left(\frat{x^2}{\rho^2}-2\right)
\ln\left(1+\frat{x^2}{\rho^2}\right)\nonumber \\ [-.2cm]
\\[-.25cm]
&-&
\left.
\frat{x^2}{\rho^2}\left(\frat{x^2}{\rho^2}-2\right)
\frat{\sqrt{c}}{\rho^2}\ln\frat{\rho^2(\sqrt{c}-x^2)}
{(\rho^2+x^2)\sqrt{c}+x^4+3\rho^2x^2}\right\}~,\nonumber
\end{eqnarray}
where the notation $c=x^4 + 4\rho^2 x^2$ was used again.
The asymptotic behaviour of these functions is
\begin{equation}
\label{SecOrd_asj1}
\lim_{x \to 0}  J_1(x,\rho) \to \pi^2\rho^2~, \qquad
\lim_{x \to \infty} J_1(x,\rho) \to 2 \pi^2\frat{\rho^4}{x^2}~.
\end{equation}
\begin{equation}
\label{SecOrd_asj2}
\lim_{x \to 0}  J_2(x,\rho) \to \frat{2\pi^2}{3}~
\ln (x^2/\rho^2)~, \qquad
\lim_{x \to \infty} J_2(x,\rho) \to -4\pi^2\frat{\rho^4}{x^4}~.
\end{equation}
The solution of the system of equations
(\ref{SecOrd_0}) can be written as
\begin{eqnarray}
\label{Fit1}
D^{(2)}(x)&=&9\pi^4~\frat{N_c}{N_c^{2}-1}~
\frat{(n_4+\bar n_4)^2~\rho^4}{2} I^{(2)}(x/\rho)~,
\nonumber\\ [-.2cm]
\\ [-.25cm]
 D^{(2)}_1(x)&=&-9\pi^4~\frat{N_c}{N_c^{2}-1}
 ~\frat{(n_4+\bar n_4)^2\rho^4}{2} I^{(2)}_1(x/\rho)~.
 \nonumber
\end{eqnarray}
where the functions $I^{(2)},~I_1^{(2)}$
have been determined numerically.
These functions turn out
to be non-negative for all $x$.
Their values at zero distance
are $~I^{(2)}(0)=1.28$ and $~I_1^{(2)}(0)=0.28$. Taking the
first and second order terms in the density expansion into account,
Eqs. (\ref{eq61}) and (\ref{Fit1}), one obtains the following form of
the correlation functions in the dilute $I\bar{I}$ gas model
\begin{eqnarray}
\label{Fit2}
D(x)&=&8(n_4+\bar n_4)~I(x/\rho)+
~9\pi^4\frat{N_c}{N_c^{2}-1}~\frat{(n_4+\bar n_4)^2\rho^{4}}{2}
~I^{(2)}(x/\rho)~,
\nonumber\\ [-.2cm]
\\ [-.25cm]
D_1(x)&=&-9\pi^4~\frat{N_c}{N_c^{2}-1}
~\frat{(n_4+\bar n_4)^2\rho^{4}}{2}~I_1^{(2)}(x/\rho)~\,
~,\nonumber
\end{eqnarray}
where $I(x/\rho)$ denotes one of the possible parametrizations of the
first order contribution (for physical reasons preferably $I_{ph}(x/\rho)$). 
We stress again that in agreement with the
lemma above the violation of (anti-)selfduality,
not only in instanton-anti-instanton but also
in instanton-instanton and anti-instanton-anti-instanton superpositions,
leads to the contributions $\propto (n_4+\bar n_4)^2$.
Besides these approximative solutions, there are also exact 
multiinstanton or multi-anti-instanton 
solutions of the Euclidean field equations.
They are suppressed by higher action, and
the corresponding distribution of their collective coordinates is even less
certain within the real vacuum. Fortunately, they cannot give a contribution
to $D_1$ due to their strict (anti-)selfduality. A comparison between the
leading and second order (in the density) contributions to $~D~$ leads to
an upper bound for the packing fraction
$n_4^t~\rho^4 \ll \frat{2^3}{3^3\pi^2}~\frat{N_c^2-1}{N_c}$.
As long as this holds, the density
expansion of the dilute gas approximation should be reliable.

With our numerical solution for $I_1^{(2)}$ one can see from (\ref{Fit2})
that the instanton contribution to $D_1(x)$ is negative for all $x,$ 
in contrast to the non-perturbative contribution as extracted from best
fits to lattice data in \cite{dig1}--\cite{Bali}.
This fact rises several questions. First of all, one might wonder
whether the existing lattice data are incompatible with a
negative non-perturbative contribution to $D_1$,
and hence are in contradiction with the
instanton-gas model. Our experience with the existing lattice data
tells, that the fits are not conclusive concerning the sign of
the pure exponential contribution in $D_1$.
Other reasonable fits are possible as well.
One such solution
found by an own fit to the lattice data of Ref. \cite{dig2}
is shown in Fig. 2.
The corresponding parameters in (\ref{eq5}) are
\begin{eqnarray}
\label{eq6c}
\frat{A}{\Lambda^{4}_L} \approx \xmn{2.52}{8}~, \quad
\frat{B}{\Lambda^{4}_L} \approx \xmn{-5.14}{7}~, &\quad&
a \approx  0.415~,  \quad  b \approx  0.307~, \nonumber \\
\lambda_A \approx 0.22 \mbox{fm} ~, \quad \lambda_B \approx 0.24~{\mbox fm},
&\quad&
\lambda_a = \lambda_b \to \infty~.
\end{eqnarray}
This fit achieves a reasonable value $\chi^2/N_{d.f.}\simeq 3.45$.

A second question directly addresses the phenomenological consequences of
negative $D_1$. It was shown in \cite{simdosch} that $D$ and $D_1$
define the scalar and spin-dependent potentials of heavy quarkonia.
In \cite{Bad} functions $D$ and $D_1$ of Gaussian 
shape have been used to predict
the spin-orbit and hyperfine splitting of charmonium and bottomonium,
and it was shown that only negative $D_1$ can reproduce the experimental
situation. Therefore, it is quite possible that $D$ and $D_1$, as obtained
from the instanton gas model, can generate a phenomenologically
reasonable spectrum of heavy quarkonia. Work in this direction is
planned for the future.

The resulting combined fit of both $D$ and $D_1$ with our model dependent
input functions (using $I_{ph}$)
is shown in Fig. 3 
together with the fitted exponential contributions to 
the lattice data for quenched QCD.  
One can see that it is still possible to imitate lattice data with $D$
and $D_1$ constrained by the instanton model.
The result of the best fit
including the second order correction is the following:
\\

\begin{center}
\begin{tabular}{|c|c|c|c|} \hline
  quenched QCD             &   $I_r$    &  $I_{ph}$  & $I_s$ \\ \hline
$\rho/\lambda_A$           & 0.78     &  1.29      & 1.85\\
$n^{t}_4~/\mbox{fm}^{-4}$  & 4.76     &  3.42      & 2.97\\ \hline
\end{tabular}
\end{center}

One should note that the shape of $I,~I^{(2)},~I_1^{(2)}$
as functions of $x/\rho$ depends on the path taken for the Schwinger phase
factor and on the profile of individual instantons.
Concerning the choice of the path the straight line path (corresponding to
$I_{ph}$) is the most natural one in view of the way the correlators
are measured on the lattice.
As far as the profile is concerned, in the present paper we have chosen 
the exact
classical solution (\ref{eq51/52}), while its shape is expected to
be changed due to the
interactions between the (anti-)instanton and the vacuum (medium)
on the classical and quantum level.
Therefore, we can choose a more general instanton potential 
\begin{eqnarray}
\label{One}
a_\mu^{a}(x)  =  2 \,\bar\eta_{a \mu \nu}\,
\frat{x_\nu}{x^2} f(x)~,
\end{eqnarray}
as proposed in \cite{diak}
with a profile function $~f$ behaving for large $x$  like $\sim \exp(-mx)$,
with $m^2=\frac{27}{8}~\pi^2~(n_4+\bar n_4)~\rho^2.$
This will certainly allow to describe the lattice results in a better way.
The numerical evaluation of this effect is now in progress.

\section{Conclusions and Discussion}
\label{sect.5}

We have considered the semiclassical approximation for the non-Abelian
field correlators. The dilute instanton gas model was used.
Generically, the correlators for an individual background configuration
had to be evaluated numerically, partly in order to take into account
the Schwinger line correctly.

Let us briefly summarize our main results. The comparison of the 
considered correlators with lattice data at zero temperature
 \cite{dig2}, \cite{dig4},
shows the following. The nonperturbative part of the
correlator $~D$  can be reasonably described by the mixed
$I~-~\bar I$ gas (see Fig. 3).
It was shown analytically that selfdual configurations contribute only
to the correlators $D$ and $D_2$
in the leading order of the density expansion, while $D_1$ appears
only in the second order. This  may explain qualitatively the large
ratio $D/D_1$ found in lattice simulations \cite{dig1,dig2,dig3,dig4}.

Another interesting result is that non-positivity of $D_1$ is characteristic
for the instanton gas model. We have demonstrated that this feature
is compatible with lattice data. Moreover, we have argued that this
might be even favourable from a phenomenological point of view.

The fitted values of the instanton densities turn out to be large
in the case of quenched simulations (several instantons
per $1~\mbox{fm}^4$). To describe this case consistently within
the instanton density expansion a medium correction for the instanton profile
\cite{diak}
must be taken into account. In the case of full QCD (with dynamical fermions)
the situation is more safe for the naive instanton superposition.
For this case, the estimated values for instanton density and radius
are found in the right ballpark known from instanton phenomenology, as well
as the estimated gluon and quark condensate values extracted from the lattice
data were consistent with QCD sumrules.
 
In this paper the effect of the inclusion of parallel transporters 
(Schwinger lines $S$) into a semiclassical
calculation (and the path dependence) has been studied for the first time.
It is displayed in Fig. 1 that the effect
(compared with a choice of path with $S=1$) 
is of the order of $20$ \% in the region of physical interest.

We did not discuss here the question whether instantons,
which yield phenomenologically realistic hadron correlators \cite{shur2}
give a complete description of the  QCD vacuum configurations. The point is
that the dilute instanton gas model (also in the way we have treated it here)
does not confine. Within the vacuum correlator model, the nonzero contribution
to the string tension which would be obtained from our correlator $D$, would
be cancelled by higher correlators (see \cite{sim} for discussion and
more references). Therefore to get realistic vacuum there are at least
two possibilities: to modify the instanton model in a way that instantons give 
only contributions of Gaussian nature (some Gaussian instanton ensemble)
or to assume additional contributions, like dyons and antidyons. Work
in this second direction is now in progress.

\par\vspace*{1cm}\noindent
{\bf Acknowledgements}

The financial support of RFFI, grants 96-02-16303, 96-02-00088G
and 97-02-17491,
and through the joint RFFI-DFG project 436 RUS 113/309/0 (R)
is gratefully acknowledged. E.-M. I. has been supported by the DFG
under grant Mu932/1-4.

This investigation has begun while one of the authors (S.~V.~M.) was
visiting the Institut f\"ur Physik of the Humboldt-Universit\"at zu
Berlin. It is a pleasure  for him to thank all the staff of the Institute
for kind hospitality.  B.~V.~M. thanks
the Institut f\"ur Physik of the Humboldt-Universit\"at zu Berlin for the
hospitality and financial support while this work has been proceeded.
E.-M.~I. gratefully acknowledges the hospitality of ITEP Moscow and the
travel grant offered by the DFG.
The authors are grateful to A.M. Badalian, H.G. Dosch and A.Di Giacomo
for enlightening discussions.
Last but not least they express their gratitude to A. Di Diacomo and his
collaborators for providing them with the table of the lattice data
of Ref. \cite{dig2}.


\newpage

\noindent
{\large FIGURE CAPTIONS: }

\vspace{1cm}

\noindent
FIG. 1:
The one-instanton contributions $~I_{r,ph,s}$
to the correlator $~D$, according to Eq. (\ref{rep}),
for different paths in the Schwinger line factor $~S$, Eq. (\ref{eq9}).
The upper solid curve refers to $~S=1$ for the regular gauge (anti-)instanton
($~I_{r}$), the dashed line to the straight-line path ($~I_{ph}$),
and the lower solid line to (anti-)instantons in the singular gauge
with the infinite arc path as explained in the text ($~I_{s}$).

\noindent
FIG. 2:
Fits to $~D_{\bot}$ (upper curves) and
$~D_{||}$ (lower curves), as defined in (\ref{par/perp}), given in units of 
$(\mbox{lattice spacing} \cdot \Lambda_L)^4$ versus physical 
distance (in units of fermi). The lattice data are taken from
Ref. \cite{dig2}. Solid lines represent the fit
according to the parametrization (\ref{eq5},\ref{eq6b}).
Dashed lines show our own fit  (\ref{eq6c})
with a negative $~D_1,$ as discussed in the text. 

\noindent
FIG. 3:
The correlator $~D$ with the second order corrections
(\ref{Fit2}) taken into account (upper solid line)
and the correlator $~|D_1|$ (lower solid line) as given by the
instanton gas model compared with the non-perturbative parts extracted
from lattice data for quenched QCD \cite{dig2}
(fit (\ref{eq6c}), upper and lower dashed line, respectively).
For comparison, the long-dashed line shows the first
order density contribution to $~D$  with the  
straight Schwinger line expression $~I_{ph}$.
The evaluated curves correspond to the parameters
$n^t_4=3.42 ~\mbox{fm}^{-4},~\rho=1.29\lambda_A$. \\

\newpage
\begin{figure}
\label{Fig1}
\begin{center}
\grpicture{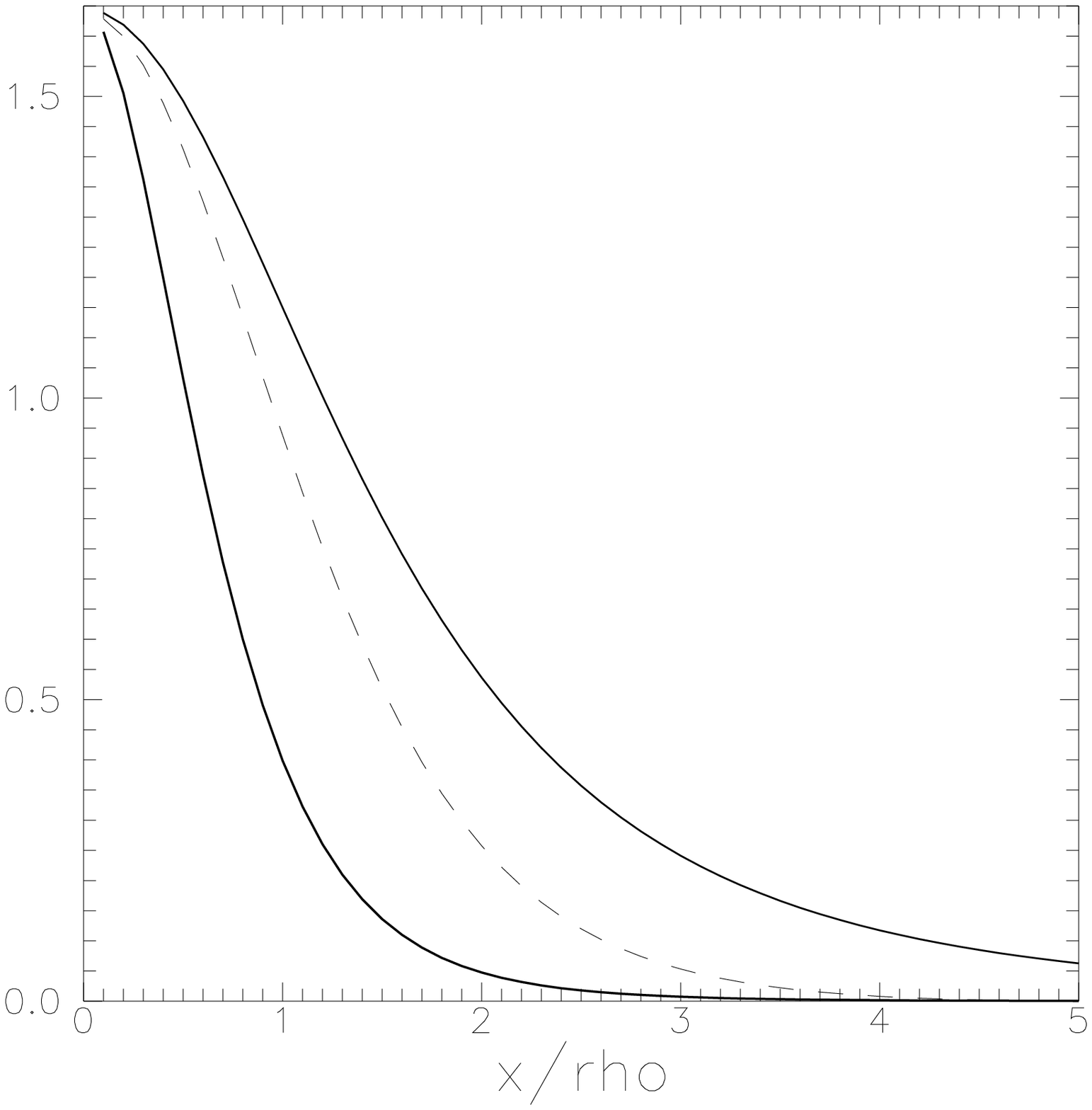}
\vspace*{2cm}
\caption{ }
\end{center}
\end{figure}

\newpage
\begin{figure}
\label{Fig2}
\begin{center}
\grpicture{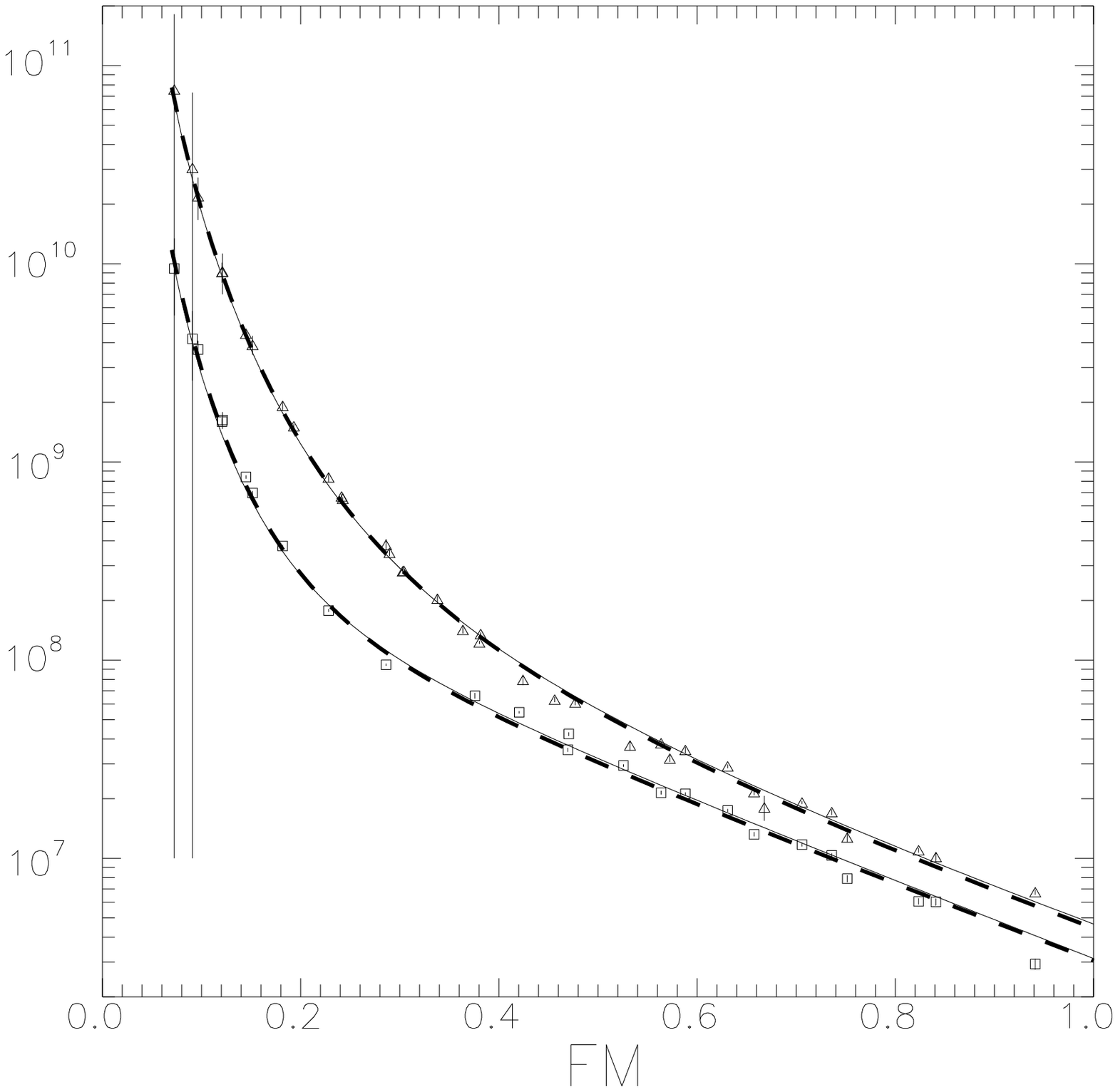}
\vspace*{2cm}
\caption{ }
\end{center}
\end{figure}

\newpage
\begin{figure}
\label{Fig3}
\begin{center}
\grpicture{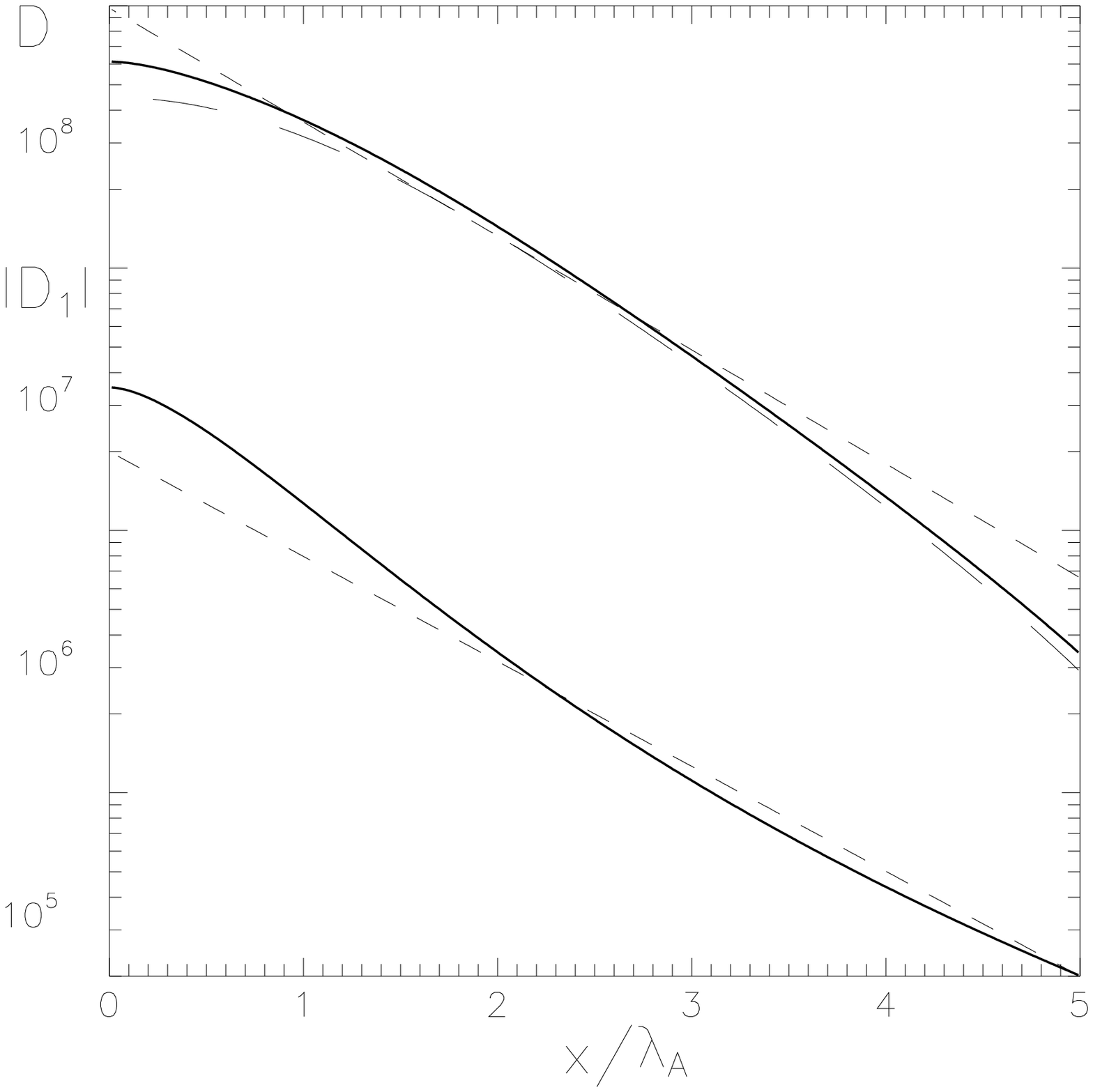}
\vspace*{2cm}
\caption{ }
\end{center}
\end{figure}

\end{document}